\documentclass[10pt,aps,pra,tightenlines,nofootinbib,onecolumn]{revtex4-2}

\usepackage[T1]{fontenc}
\usepackage[margin=1.5cm]{geometry}
\usepackage{amsmath}
\usepackage{amssymb}
\usepackage{amsthm}
\usepackage{dsfont}
\usepackage{mathtools}
\usepackage[normalem]{ulem}
\usepackage{mathrsfs}
\usepackage[makeroom]{cancel}
\usepackage{graphicx}
\usepackage{bm}
\usepackage{datetime}
\usepackage{physics}
\usepackage[dvipsnames]{xcolor}
\usepackage{comment}
\usepackage{wasysym}
\usepackage{booktabs}

\usepackage[colorlinks=true,
            urlcolor=blue,
            linkcolor=blue,
            citecolor=blue]{hyperref}

\newtheorem{thm}{Theorem}

\newtheorem{defn}{Definition}

\newtheorem*{lem*}{Lemma}
\newtheorem*{thm*}{Theorem}

\def\ii{\mathrm{i}}
\def\e{\mathrm{e}}

\def\id{\mathds{1}}

\newcommand{\jacobi}[2]{\left(\frac{#1}{#2}\right)}

\makeatletter
\def\Dated@name{}
\makeatother

\begin{document}

\allowdisplaybreaks

\title{Maximally Entangled Mutually Unbiased Bases \\
in Bipartite Systems of Unequal Prime Dimensions}

\author{Bogdan S. Damski$^1$}
\author{Karol \.{Z}yczkowski$^{1,2}$}

\affiliation{
$^1$Jagiellonian University,
Faculty of Physics, Astronomy and Applied Computer Science,
{\L}ojasiewicza 11, 30-348 Krak\'ow, Poland
}

\affiliation{
$^2$Center for Theoretical Physics, Polish Academy of Sciences, Warsaw
}

\date{November 13, 2026}

\begin{abstract}
Mutually unbiased bases with prescribed entanglement properties are of fundamental interest in quantum information theory.
In the interesting case of a bipartite Hilbert space of unequal prime dimensions,
$\mathcal{H}_p\otimes\mathcal{H}_q$, with $p<q$ prime, the standard solution provides $p+1$ product bases.
For any such dimension $d=pq$ we construct $p+1$ bases, such that $p$ of them are represented by complex Hadamard matrices, each equivalent to a Fourier matrix $F_d$, with all vectors maximally entangled.
Our explicit construction is based on eigenvectors of suitable Weyl--Heisenberg operators, the Chinese Remainder Theorem and quadratic Gaussian sums.
The obtained result establishes a lower bound for the number of maximally entangled mutually unbiased bases, $M(p,q)\geq p$.

\end{abstract}

\maketitle

\section{Introduction}

Mutually unbiased bases (MUBs) play a fundamental role in quantum information theory and have attracted considerable attention due to both their theoretical significance and their practical applications. The existence of complete sets of MUBs in arbitrary dimensions remains a major open problem in quantum information science. The importance of this problem is reflected in its inclusion among the ``Five open problems in quantum information'' \cite{Horodecki2022}.

The existence of a complete set of MUBs remains unresolved even in dimension six. This problem is closely related to a conjecture formulated by Zauner in his seminal work on quantum designs \cite{Zauner2011}. Specifically, Zauner conjectured that no more than three MUBs can exist in dimension six. This conjecture is widely believed to hold and is supported by extensive numerical evidence \cite{Butterley2007,Raynal2011,Matolcsi2015,Matolcsi2026} and more recent AI-aided analytical results \cite{Wuttig2026_1, OpenAI2026_1, OpenAI2026_2,Wuttig2026_2}.

Beyond their fundamental significance, MUBs have found applications in a wide range of areas in quantum information science. In particular, they have been used in quantum state tomography and estimation \cite{Ivanovic1981,Wootters1989,Adamson2010,Lima2011,Giovannini2013}, quantum key distribution \cite{Bennett2014,Bechmann_Pasquinucci2000,Cerf2002,Yu2008,Bouchard2018,Mafu2013,Mirhosseini2015}, entanglement detection \cite{Spengler2012,Kalev2013,Chen2014,Graydon2016,Bae2019,Hiesmayr2021,Farkas2021,Ikuta2022}, and quantum steering \cite{Wittmann2012,Kalev2014,Cavalcanti2016,Zeng2018,Huang2021}.

Consider a composite bipartite system
$\mathcal{H}_{d}\otimes\mathcal{H}_{d'}$, with $d\leq d'$.
If both subsystems admit at least $d+1$ MUBs, then $d+1$ MUBs
can be constructed in the composite space by taking tensor
products of the corresponding local bases \cite{McNulty2026}. Although this provides a natural construction of MUBs in composite systems, extending it to a complete set of
$dd'+1$ MUBs requires additional constraints. In particular, for a
complete set of MUBs, the average purity of the reduced states is
given by Lubkin's formula \cite{Lubkin1978},
\begin{equation}
\Bigl\langle \Tr\sigma^2 \Bigr\rangle
=
\Bigl\langle \Tr\sigma'^2 \Bigr\rangle
=
\frac{d+d'}{1+dd'}.
\end{equation}
where $\sigma$ and $\sigma'$ denote the reduced density matrices on
$\mathcal{H}_{d}$ and $\mathcal{H}_{d'}$, respectively. If a complete set contains exactly $d+1$ product bases, then
all the remaining bases must consist entirely of maximally
entangled states. This follows from Lubkin's formula, since
the average purity of the reduced states associated with the
remaining bases must attain its minimum possible value, $1/d$.
This observation motivates the study of so-called
\emph{maximally entangled MUBs}, that is, sets of MUBs in
$\mathcal{H}_{d}\otimes\mathcal{H}_{d'}$ in which all states
forming the constituent bases have entanglement entropy equal
to $\log d$. In particular, one is interested in determining
the maximum number of such bases that can coexist in
$\mathcal{H}_{d}\otimes\mathcal{H}_{d'}$.

One of the well-known approaches to constructing maximally
entangled MUBs in $\mathcal{H}_{d}^{\otimes 2}$ is based on
classical Latin squares. It relies on identifying a pair of
mutually orthogonal Latin squares $\{L,L'\}$ of order $d$,
together with a Hadamard matrix $H$ of the same order
\cite{Wocjan2004}. A related construction was later developed
using quantum Latin squares. In this approach, one identifies
a pair of weakly orthogonal quantum Latin squares $\{Q,Q'\}$
of order $d$, together with a pair of Hadamard matrices
$\{H,H'\}$ of the same order \cite{Musto2017}. The concepts
of classical and quantum Latin squares, as well as their
properties, are discussed in
\cite{Wocjan2004, Musto2017, McNulty2026}.
Both approaches can be extended to larger collections of maximally
entangled MUBs, provided that the corresponding conditions are
satisfied pairwise for every pair of bases in the collection.
However, a significant limitation shared by these constructions is the need to identify suitable Latin squares, quantum Latin squares, and Hadamard matrices on a case-by-case basis.
This makes it difficult to derive closed-form constructions
applicable to arbitrary dimension $d$. More generally, existing constructions of maximally entangled MUBs, both in $\mathcal{H}_{d}^{\otimes 2}$
and in higher-dimensional bipartite systems
$\mathcal{H}_{d}\otimes\mathcal{H}_{d'}$, often rely on
 ad-hoc dimension-dependent methods. Various approaches to
constructing maximally entangled bases in
$\mathcal{H}_{d}\otimes\mathcal{H}_{d'}$, together with their
connections to MUBs, have been investigated in
\cite{Tao2015,Luo2016,Xu2017,Cheng2017,Liu2017,Xu2019,Zang2022},
with some of these methods reviewed in \cite{Shi2020}.

Nevertheless, several lower bounds on the number $M(d,d')$ of constructible
maximally entangled MUBs in $\mathcal{H}_{d}\otimes\mathcal{H}_{d'}$
are known for various classes of dimensions. For example, let
$d=\prod_i p_i^{n_i}$ and $k=\prod_i q_i^{m_i}$ be the prime
factorizations of $d$ and $k$, where $p_i$ and $q_i$
are prime numbers while $n_i$ and $m_i$ are positive integers. Then, we have
\cite{Cheng2017,Liu2017,Xu2017,Shi2020}
\begin{equation}
M(d,d)\geq
\frac{3+(-1)^{d+1}}{2}\min_i(p_i^{n_i}-1),
\
M(d,kd)\geq
\min\left\{\min_i(q_i^{m_i}+1),M(d,d)\right\}.
\end{equation}
The aim of the present work is to provide a closed-form
construction of $p$ maximally entangled MUBs in
$\mathcal{H}_{p}\otimes\mathcal{H}_{q}$, where $p$ and $q$ are
distinct primes such that $p<q$. Our construction establishes the lower bound
$M(p,q)\geq p$ for arbitrary distinct primes $p$ and $q$,
thereby extending previous results for specific pairs of
dimensions to a general class of bipartite systems.

The preceding discussion focuses on MUBs whose basis states are
maximally entangled with respect to a fixed bipartition. This
condition can be viewed as a special case of a broader notion of
isoentanglement, in which all basis states have the same degree of
entanglement, not necessarily maximal, with respect to any
prescribed bipartition of the system. Sets of MUBs satisfying
this property are referred to as isoentangled MUBs. A complete
set of five isoentangled MUBs is known in
$\mathcal{H}_2^{\otimes 2}$ \cite{Czartowski2020}, whereas the
existence of complete sets of isoentangled MUBs in
$\mathcal{H}_p^{\otimes 2}$ remains an open problem for prime
$p>2$. A related notion was recently introduced for mutually unbiased
Hadamard matrices (MUHMs) \cite{Damski2026}. Isoentangled MUHMs
are sets of MUHMs whose column vectors share the same degree
of entanglement with respect to any prescribed bipartition of the
system. In dimension $d$, a complete set consists of $d$ MUHMs.
Complete sets of isoentangled MUHMs were constructed in
$\mathcal{H}_p^{\otimes n}$ for prime $p>2$ \cite{Damski2026}.
Beyond their fundamental interest, isoentangled bases may also
be relevant to practical applications in quantum information
processing, particularly in joint measurements and quantum
networks \cite{DelSanto2024}.

This paper is organized as follows. In Sec. \ref{sec:setting}, we introduce the basic concepts and notation used throughout the paper, including MUBs, Hadamard matrices and Weyl--Heisenberg operators. Sec. \ref{sec:maximally_entangled_MUBs} contains the main result of this work: the derivation of a closed-form construction of $p$ maximally entangled MUBs in $\mathcal{H}_p\otimes\mathcal{H}_q$, where $p$ and $q$ are distinct prime numbers satisfying $p<q$. Finally, in Sec. \ref{sec:summary}, we discuss the implications of our results.

\section{Setting the Scene}
\label{sec:setting}

In this section, we review the basic concepts used throughout the paper. We begin by introducing mutually unbiased bases (MUBs), mutually unbiased Hadamard matrices (MUHMs) and the Weyl--Heisenberg operators.

\begin{defn}[Mutually unbiased bases]
We say that two orthonormal bases
$\mathcal{B}_1=\{\ket{e_i}\}_{i=1}^d$ and
$\mathcal{B}_2=\{\ket{f_j}\}_{j=1}^d$ in $\mathcal{H}_d$ are
\emph{unbiased} if the squared overlap between any two states belonging to
different bases is constant,
\begin{equation}
\left|\braket{e_i}{f_j}\right|^2=\frac{1}{d}.
\label{eq:MUB}
\end{equation}
A collection of $N$ orthonormal bases
$\{\mathcal{B}_k\}_{k=1}^N$ is said to be
mutually unbiased if every pair of distinct bases in the collection
is unbiased.
\end{defn}

The defining relation of MUBs is invariant under a simultaneous
unitary transformation of all basis vectors. Hence, if a collection
of MUBs is subjected to the same unitary transformation, the resulting
collection also forms a set of MUBs that is unitarily equivalent to
the original one. Consequently, one of the bases can always be chosen
as the computational basis. In this representation, the vectors of
any basis unbiased with respect to the computational basis form the
columns of a complex Hadamard matrix. More generally, a complete set of $d+1$ MUBs is unitarily equivalent
to a collection consisting of the identity operator
$\id_d\in \mathcal{U}(d)$ and $d$ pairwise mutually unbiased
complex Hadamard matrices
$\{H_i\in \mathcal{U}(d)\}_{i=1}^d$
\cite{Durt2010}. Throughout this manuscript, $\mathcal{U}(d)$ denotes
the space of unitary operators acting on $\mathcal{H}_d$. 

\begin{defn}
A matrix $H\in \mathcal{U}(d)$ is called a complex Hadamard matrix
if its entries satisfy
\begin{equation}
|H_{ij}|=\frac{1}{\sqrt{d}}, \ HH^\dagger=\id_d.
\end{equation}
We distinguish the following properties.
\begin{enumerate}
    \renewcommand{\labelenumi}{(\roman{enumi})}

    \item $H\in \mathcal{U}(d)$ is said to be of Butson type
    $\mathrm{BH}(d,\alpha)$, up to a global phase, if its entries satisfy
    \begin{equation}
    H_{ij}\sim \frac{1}{\sqrt{d}}\omega_\alpha^{k_{ij}},
    \end{equation}
    where $k_{ij}\in\mathbb{Z}_\alpha$,
    $\omega_\alpha=\e^{2\pi \ii/\alpha}$, and $\sim$ denotes equality
    up to a common global phase. More precisely, for matrices $A$ and $B$, $A\sim B$ means
    that $A=\e^{\ii\theta}B$ for some $\theta\in\mathbb{R}$.

    \item Two complex Hadamard matrices
    $H_1,H_2\in \mathcal{U}(d)$ are called mutually unbiased if
    \begin{equation}
    \left|H_1^\dagger H_2\right|
    =\frac{1}{\sqrt{d}}.
    \end{equation}
    A collection of $N$ Hadamard matrices
    $\{H_k\in\mathcal{U}(d)\}_{k=1}^N$ is said to be
    mutually unbiased if every pair of distinct Hadamard matrices
    in the collection is unbiased.

    \item Two complex Hadamard matrices
    $H,H'\in \mathcal{U}(d)$ are called equivalent if
    there exist diagonal unitary matrices
    $D,D'\in \mathcal{U}(d)$ and permutation matrices
    $P,P'\in \mathcal{U}(d)$ such that
    \begin{equation}
    H'=PDHD'P'.
    \end{equation}
\end{enumerate}
\end{defn}

An important property of complex Hadamard matrices is that Butson-type
Hadamard matrices $\mathrm{BH}(d,d)$ exist for every positive integer
$d$. An explicit construction is provided by the Fourier operator
$F_d\in \mathcal{U}(d)$, defined with respect to the computational
basis as
\begin{equation}
    F_d\ket{j}
    = \frac{1}{\sqrt{d}}
    \sum_{k=0}^{d-1}
    \omega_d^{-jk}\ket{k}.
\end{equation}
 More generally, the existence of Butson-type Hadamard matrices $\mathrm{BH}(d,\alpha)$ depends on the choice of the parameters $d$ and $\alpha$. In particular, if $\beta$ is an odd prime, a necessary condition for the existence of $\mathrm{BH}(d,\beta)$ is that $d=m\beta$ for some positive integer $m$ \cite{Butson1962}. Moreover, it is conjectured that such matrices exist for all these dimensions \cite{Beder1998}. For $m=2$, their existence can be established by an explicit construction \cite{Butson1962}. Nevertheless, a complete characterization of the parameter pairs for which Butson-type Hadamard matrices exist remains an open problem \cite{Heng1998,deLauney1984}. Beyond the existence problem, complex Hadamard matrices with various interesting algebraic and structural properties have also attracted considerable attention in the literature (see e.g. \cite{Bruzda2018,Bruzda2023,Bruzda2024_1,Bruzda2024_2}).

Complete sets of $d+1$ MUBs are known to exist in prime dimensions $d=p$ \cite{McNulty2026}. One such set can be constructed from the eigenbases of the Weyl--Heisenberg operators $\{Z\}\cup\{XZ^k\}_{k=0}^{p-1}$ \cite{Bandyopadhyay2001}. These operators are built from the shift and clock operators, defined as follows.

\begin{defn}[Shift and clock operators]
Consider the shift and clock operators $X,Z\in \mathcal{U}(d)$, defined by their action on the computational basis as \cite{Weyl1927}
\begin{equation}
X\ket{i}=\ket{i+1 \bmod d}, \ Z\ket{i}=\omega_d^i\ket{i}.
\end{equation}
The operators
\begin{equation}
W_{kl}=\tau_d^{kl}X^kZ^l, 
\end{equation}
where $\tau_d=-\e^{\ii\pi/d}$ and $k,l\in\mathbb{Z}_d$ are called Weyl--Heisenberg operators. 
\end{defn}

To make the prime-dimensional construction of MUBs explicit,
we recall the eigenvectors of the Weyl--Heisenberg
operators in prime dimensions. Below we present a convenient form
of the standard solution \cite{Bandyopadhyay2001}
of $(p+1)$ MUBs in a prime dimension $p$,
which will be used in the construction
developed in Sec. \ref{sec:maximally_entangled_MUBs}.

\begin{thm}\label{thm:eigenvectors}
Let $p$ be a prime number. Then, the following statements hold.

\begin{enumerate}
    \item[(i)] The computational basis
    \begin{equation}
    \left\{\ket{i}\right\}_{i=0}^{p-1}
    \label{eq:WH_eigenvectors_a0}
    \end{equation}
    is an eigenbasis of $Z^r\in \mathcal{U}(p)$
    for any $r\in\mathbb{Z}$.

    \item[(ii)] For each $r\in\mathbb{Z}$, the eigenvectors of
    $XZ^r\in \mathcal{U}(p)$ form a basis given by
    \begin{equation}
    \mathcal{B}_r^{(p)}=
    \Bigg\{
    \ket{\phi_p(r,t)}
    =
    \frac{1}{\sqrt{p}}
    \sum_{j=0}^{p-1}
    \tau_p^{j[r(j-p)-2t]}\ket{j}
    \Bigg\}_{t=0}^{p-1}.
    \label{eq:v_p}
    \end{equation}

    \item[(iii)] The eigenvector bases of the Weyl--Heisenberg operators
    $\{XZ^r\}_{r=0}^{p-1}\cup\{Z\}$
    constitute a complete set of $p+1$ MUBs in $\mathcal{H}_p$.
    Moreover, the matrix
    \begin{equation}
    H_r^{(p)}=
    \bigl(
    \ket{\phi_p(r,0)}\,
    \ket{\phi_p(r,1)}\,
    \cdots\,
    \ket{\phi_p(r,p-1)}
    \bigr)\in \mathcal{U}(p)
    \end{equation}
    is a complex Hadamard matrix equivalent to the Fourier matrix
    $F_p$.
\end{enumerate}
\end{thm}

\begin{proof}
The statement (i) follows directly from
\begin{equation}
Z^r\ket{i}=\omega_p^{ri}\ket{i}.
\end{equation}
In order to prove statement (ii), we compute
\begin{equation}
\begin{aligned}
XZ^r\ket{\phi_p(r,t)}
&=
\frac{1}{\sqrt{p}}
\sum_{j=0}^{p-1}
\tau_p^{j[r(j-p)-2t]+2jr}\ket{j+1}\\
&=
\frac{1}{\sqrt{p}}
\sum_{j=0}^{p-1}
\tau_p^{(j-1)[r(j-1-p)-2t]+2(j-1)r}\ket{j}\\
&=
\tau_p^{2t+r(p-1)}
\frac{1}{\sqrt{p}}
\sum_{j=0}^{p-1}
\tau_p^{j[r(j-p)-2t]}\ket{j}\\
&=
\tau_p^{2t+r(p-1)}
\ket{\phi_p(r,t)}.
\end{aligned}
\end{equation}
Hence, $\ket{\phi_p(r,t)}$ is an eigenvector of $XZ^r$
to eigenvalue $\tau_p^{2t+r(p-1)}$. This observation 
implies (ii). This representation of the eigenvectors
is equivalent to the one given in \cite{Bandyopadhyay2001}, but
is more convenient for the calculations that follow in the rest of this work.

We now prove statement (iii). Let $D_r\in \mathcal{U}(p)$ be the diagonal operator defined by its
action on the computational basis as
\begin{equation}
    D_r\ket{j}
    = \tau_p^{rj^2-2rj\delta_{p,2}}\ket{j},
\end{equation}
where $\delta_{i,j}$ is the Kronecker delta. By direct computation we obtain
\begin{equation}
\begin{aligned}
D_rF_p\ket{t}
&=D_r\left(
\frac{1}{\sqrt{p}}
\sum_{k=0}^{p-1}\omega_p^{-kt}\ket{k}
\right)\\
&=\frac{1-\delta_{p,2}}{\sqrt{p}}
\sum_{k=0}^{p-1}
\tau_p^{rk^2}\omega_p^{-kt}\ket{k}
+\frac{\delta_{p,2}}{\sqrt{p}}
\sum_{k=0}^{p-1}
\tau_p^{rk^2-2rk}\omega_p^{-kt}\ket{k}=\ket{\phi_p(r,t)}.
\end{aligned}
\end{equation}
This implies that $H_r^{(p)}=D_rF_p$. Therefore, $H_r^{(p)}$ is a complex
Hadamard matrix equivalent to $F_p$ as stated in (iii).
\end{proof}

For positive integer $n>1$, the construction of a complete set of $d+1$ MUBs in dimensions
$d=p^n$ is more involved and relies on finite fields
$\mathbb{F}_{p^n}$ for odd primes $p$ \cite{Wootters1989}, or on Galois rings in certain
constructions involving prime-power dimensions, such as
$\mathrm{GR}\left(4,n\right)$ \cite{Klappenecker2003} for $p=2$ and $\mathrm{GR}\left(9,n\right)$ for $p=3$
\cite{Damski2026}. The situation is more complicated
in general composite dimensions $d=\prod_{i=1}^{r}p_i^{n_i}$,
where $p_1, p_2,\ldots,p_r$ are distinct prime factors of $d$,
$n_1, n_2,\ldots,n_r$ are positive integers,
and $r$ is the number of distinct prime factors. The existence and
construction of complete sets of MUBs in such dimensions remain unresolved in general.
It has been conjectured that, for $\nu=1+\min_i p_i$,
the eigenbases of the Weyl--Heisenberg operators $\{Z\}\cup\{XZ^k\}_{k=0}^{\nu-2}$
form a set of $\nu$ unextendible MUBs, meaning that no additional basis can be found that is unbiased with respect to all bases in the set
\cite{Grassl2017}. This set is said to be weakly unextendible,
meaning that it cannot be extended by adding an eigenbasis of any
other Weyl--Heisenberg operator
\cite{Grassl2017,Aschbacher2006}. Further illustrating the complexity
of the problem, in even dimensions $d=6,10,12,14$, the eigenbases of $\{Z,X,XZ\}$
form strongly unextendible sets, meaning
that no additional vector can be mutually unbiased to all three
bases \cite{Grassl2004,Grassl2017}. In dimensions $d=9$ and $d=15$,
the eigenbases of $\{Z,X,XZ,XZ^2\}$ form sets of four
unextendible MUBs, although smaller unextendible sets consisting of
three eigenbases of Weyl--Heisenberg operators are also known
\cite{Grassl2017}.

Nevertheless, in an arbitrary composite dimension $d=\prod_{i=1}^{r}p_i^{n_i}$, it is always possible to construct $\mu=1+\min_i p_i^{n_i}$ MUBs. This can be achieved by taking tensor products of MUBs in the corresponding prime-power dimensions
\cite{McNulty2026}:

\begin{thm}\label{thm:product-mubs}
Let $d=\prod_{i=1}^{r}p_i^{n_i}$,
where $p_1,p_2,\ldots,p_r$ are distinct primes and
$n_1,n_2,\ldots,n_r$ are positive integers, and define
$\mu=1+\min_i p_i^{n_i}$.
For each $i$, let
\begin{equation}
\left\{\mathcal{B}_k^{(i)}\right\}_{k=1}^{\mu}
\end{equation}
be $\mu$ MUBs in
$\mathcal{H}_{p_i^{n_i}}$. Then, the collection
\begin{equation}
\left\{
\bigotimes_{i=1}^{r}\mathcal{B}_k^{(i)}
\right\}_{k=1}^{\mu}
\end{equation}
forms a set of $\mu$ MUBs in
$\mathcal{H}_d=\bigotimes_{i=1}^{r}\mathcal{H}_{p_i^{n_i}}$.
\end{thm}
\begin{proof}
Since a complete set of $p_i^{n_i}+1$ MUBs exists in each prime-power dimension $p_i^{n_i}$ \cite{McNulty2026}, we can choose $\mu$ MUBs in each subspace $\mathcal{H}_{p_i^{n_i}}$.
Let $\ket{\xi}=\bigotimes_{i=1}^r \ket{v_a^{(i)}}\in \bigotimes_{i=1}^{r}\mathcal{B}_a^{(i)}$ and $\ket{\zeta}=\bigotimes_{i=1}^r \ket{v_b^{(i)}}\in \bigotimes_{i=1}^{r}\mathcal{B}_b^{(i)}$, where $a\neq b$. By direct computation, we obtain
\begin{equation}
|\braket{\xi}{\zeta}|^2
=\prod_{i=1}^r
\left|\braket{v_a^{(i)}}{v_b^{(i)}}\right|^2=\prod_{i=1}^r\frac{1}{p_i^{n_i}}
=\frac{1}{d}.
\end{equation}
The second equality follows from the mutual unbiasedness of the bases $\mathcal{B}_a^{(i)}$ and $\mathcal{B}_b^{(i)}$ in $\mathcal{H}_{p_i^{n_i}}$.
Moreover, for each $k\in\{1,2,\ldots,\mu\}$, the tensor product $\bigotimes_{i=1}^r\mathcal{B}_k^{(i)}$ is an orthonormal basis of $\mathcal{H}_d$, since the tensor product of orthonormal bases is an orthonormal basis. Thus, the $\mu$ tensor product bases are pairwise mutually unbiased in $\mathcal{H}_d$. Hence, the collection $\left\{
\bigotimes_{i=1}^{r}\mathcal{B}_k^{(i)}
\right\}_{k=1}^{\mu}$
forms a set of $\mu$ MUBs in $\mathcal{H}_d$.
\end{proof}

\section{Maximally Entangled MUBs in Composite Bipartite Dimensions}\label{sec:maximally_entangled_MUBs}

In this section, we restrict our attention to bipartite Hilbert spaces
$\mathcal{H}_p\otimes\mathcal{H}_q$,
 where $p$ and $q$
are distinct primes such that $p<q$. We identify
$\mathcal{H}_{pq}\equiv\mathcal{H}_p\otimes\mathcal{H}_q$
and adopt the standard lexicographic ordering
\begin{equation}
\ket{a \bmod p}\otimes\ket{b \bmod q}
\equiv\ket{aq+b \bmod pq}.
\end{equation}
Our aim is to construct a set of $p+1$ MUBs in
$\mathcal{H}_p\otimes\mathcal{H}_q$, consisting of one
separable product basis and $p$ maximally entangled bases.
The construction proceeds in two steps. In the first step, we recall a standard set of product MUBs, denoted by
$\left\{\mathcal{B}_r^{\mathrm{sep}}\right\}_{r=0}^{p}$.
In the second step, we apply a permutation
unitary $P_{pq}$ to each basis in this set. The computational
product basis remains invariant under this transformation,
while each of the remaining $p$ bases is mapped to a
maximally entangled basis. The resulting set of $p+1$ MUBs
is therefore given by
$\left\{\mathcal{B}_r^{\mathrm{ent}}
=P_{pq}\mathcal{B}_r^{\mathrm{sep}}\right\}_{r=1}^{p}
\cup\left\{\mathcal{B}_0^{\mathrm{sep}}\right\}$,
where $P_{pq}\mathcal{B}_0^{\mathrm{sep}}
=\mathcal{B}_0^{\mathrm{sep}}$.

We now turn to the first step of the procedure, concerning
a standard set of product MUBs. The following theorem
defines this set and establishes its relevant properties.
The result follows  from Theorems \ref{thm:eigenvectors}
and \ref{thm:product-mubs}, together with the Chinese Remainder
Theorem (see Appendix A).

\begin{thm}\label{thm:startMUB}
Let $p$ and $q$ be distinct primes with $p<q$. For each
$r\in\{0,1,\ldots,p\}$, define the basis
\begin{equation}
\mathcal{B}_r^{\mathrm{sep}}
=
\left\{
\ket{\chi(r,t)}\in \mathcal{H}_p\otimes\mathcal{H}_q
\right\}_{t=0}^{pq-1},
\end{equation}
where
\begin{equation}
\ket{\chi(r,t)}
=
\delta_{r,0}\ket{t \bmod p}\otimes\ket{t \bmod q}
+
(1-\delta_{r,0})
\ket{\phi_p(r,t)}\otimes\ket{\phi_q(r,t)}.
\end{equation}
Then, the following statements hold.
\begin{enumerate}
    \item[(i)] $\{\mathcal{B}_r^{\mathrm{sep}}\}_{r=0}^{p}$ is a set of $p+1$ MUBs.
    \item[(ii)] For each
    $r\in\{1,2,\ldots,p\}$, the matrix
    \begin{equation}
    H_r^{\mathrm{sep}}
    =
    \bigl(
    \ket{\chi(r,0)}\,
    \ket{\chi(r,1)}\,
    \cdots\,
    \ket{\chi(r,pq-1)}
    \bigr)
    \in \mathcal{U}(pq)
    \end{equation}
    is equivalent to the Fourier matrix $F_p\otimes F_q$.
\end{enumerate}
\end{thm}

\begin{proof}
Firstly, we prove (i). Theorem \ref{thm:eigenvectors} establishes that
complete sets of MUBs can be constructed from the eigenvector bases
of suitable Weyl--Heisenberg operators in prime dimensions $p$ and $q$.
For each $r\in\{1,2,\ldots,p\}$, we define the corresponding
local bases as follows
\begin{equation}
\mathcal{B}_r^{(p)}
=
\left\{
\ket{\phi_p(r,t)}
\right\}_{t=0}^{p-1}, \
\mathcal{B}_r^{(q)}
=
\left\{
\ket{\phi_q(r,t)}
\right\}_{t=0}^{q-1}.
\end{equation}
By the Chinese Remainder Theorem, each pair
$(t \bmod p,t \bmod q)\in\mathbb{Z}_p\times\mathbb{Z}_q$
corresponds uniquely to an element $t\in\mathbb{Z}_{pq}$.
This implies that for $r\in\{1,2,\ldots,p\}$, we have
\begin{equation}
\mathcal{B}_r^{\mathrm{sep}}
=
\mathcal{B}_r^{(p)}\otimes\mathcal{B}_r^{(q)}.
\end{equation}
Theorem \ref{thm:product-mubs} then implies that
$\{\mathcal{B}_r^{\mathrm{sep}}\}_{r=0}^{p}$ forms a set of $p+1$ MUBs in
$\mathcal{H}_p\otimes\mathcal{H}_q$.
Thus, (i) holds.

To prove (ii), recall that by Theorem \ref{thm:eigenvectors}
for each $r\in\{1,2,\ldots,p\}$ the Hadamard matrices
$H_r^{(p)}$ and $H_r^{(q)}$ are equivalent to the Fourier
matrices $F_p$ and $F_q$, respectively. Moreover,
$H_r^{\mathrm{sep}}$ is equivalent to
$H_r^{(p)}\otimes H_r^{(q)}$ by construction.
Consequently, $H_r^{\mathrm{sep}}$ is equivalent to
$F_p\otimes F_q$. This completes the proof of (ii).
\end{proof}

We now turn to the second step of the procedure, in which we apply a
permutation $P_{pq}$ that leaves the computational
product basis invariant and maps each of the remaining
product bases to a basis of maximally entangled states.
The explicit form of $P_{pq}$ and the analysis of the resulting
bases require some preliminary notation and number-theoretic
tools, which we introduce below.

We define the quadratic Gaussian sums as
\begin{equation}
    G(a,b,c) = \sum_{n=0}^{c-1} \omega_c^{an^2+bn},
    \label{eq:gauss_def}
\end{equation}
where $a,b,c\in\mathbb{Z}$ and $c\geq 1$. For an odd prime $p$, we introduce the Legendre symbol
\begin{equation}
\left(\frac{a}{p}\right)
=
\begin{cases}
0 & a\equiv0\pmod p,\\
1 & a\not\equiv0\pmod p
\text{ and }a\equiv x^2\pmod p
\text{ for some }x\in\mathbb Z,\\
-1 & \text{otherwise},
\end{cases}
\end{equation}
and auxiliary factor
\begin{equation}
\varepsilon_p=
\begin{cases}
1 & p\equiv1\pmod4,\\
\ii & p\equiv3\pmod4.
\end{cases}
\end{equation}
For an integer $d>1$ and an integer $a$ satisfying $\gcd(a,d)=1$,
we define $\eta_d(a)$ as the multiplicative inverse of $a$ modulo $d$,
i.e.,
\begin{equation}
a\eta_d(a)\equiv 1\pmod d.
\label{eq:modular_inverse}
\end{equation}
Here, $\gcd(a,d)$ denotes the greatest common divisor of $a$ and $d$. 

In terms of the modular inverse introduced above, we are now ready to define
the permutation matrix $P_{pq}\in\mathcal{U}(pq)$, used in our
construction, by its action on the computational basis as
\begin{equation}
P_{pq}\left(\ket{a}\otimes\ket{b}\right)
=
\ket{a}\otimes
\ket{\eta_q(p)(b-a)\bmod q}.
\label{eq:keyperm}
\end{equation}
Indeed, for each fixed $a\in\mathbb{Z}_p$, the map
$b\mapsto\eta_q(p)(b-a)\bmod q$ is a permutation
of $\mathbb{Z}_q$. Hence, $P_{pq}$ permutes the
computational basis states and is therefore
indeed a permutation matrix.

The properties of quadratic Gaussian sums required in
the subsequent computations are collected in Appendix A. With the notation
and mathematical tools introduced above, we are now
ready to state the main result of this work, which provides the
desired construction of $p+1$ MUBs in ${\cal H}_{pq}$, 
such that $p$ of them are maximally entangled.

\begin{thm}\label{thm:maxMUBs}
Let $p$ and $q$ be distinct primes such that $p<q$. For each
$r\in\{1,\ldots,p\}$, define the basis
\begin{equation}
\mathcal{B}_r^{\mathrm{ent}}
=
\left\{
\ket{\Psi(r,t)}\in \mathcal{H}_p\otimes\mathcal{H}_q
\right\}_{t=0}^{pq-1},
\end{equation}
where
\begin{equation}
\ket{\Psi(r,t)}
=
\frac{1}{\sqrt{pq}}
\sum_{a=0}^{p-1}\sum_{b=0}^{q-1}
\tau_{pq}^{(p+q)\left[
r(a+pb)^2-2t(a+pb)
\right]}
\ket{a}\otimes\ket{b}.
\label{eq:simpleMUBs}
\end{equation}
Moreover, for each $r\in\{1,2,\ldots,p\}$, define the matrix
\begin{equation}
H_r^{\mathrm{ent}}=
\left(
\ket{\Psi(r,0)}\,
\ket{\Psi(r,1)}\,
\cdots\,
\ket{\Psi(r,pq-1)}
\right)=P_{pq}H_r^{\mathrm{sep}}\in \mathcal{U}(pq).
\label{eq:Hadamard_Relation}
\end{equation}

Then, the following statements hold.
\begin{enumerate}
\item[(i)] The collection $\{\mathcal{B}_0^\mathrm{sep}\}\cup\{\mathcal{B}_r^{\mathrm{ent}}\}_{r=1}^{p}$ forms a
set of $p+1$ MUBs, where $\mathcal{B}_0^\mathrm{sep}$ is the computational basis and every vector in $\mathcal{B}_r^\mathrm{ent}$ is maximally entangled for
$r\in\{1,2,\ldots,p\}$.

\item[(ii)] The collection $\{H_r^{\mathrm{ent}}\}_{r=1}^{p}$ consists of $p$
maximally entangled MUHMs,
where each $H_r^{\mathrm{ent}}$ is equivalent to the Fourier
matrix $F_p\otimes F_q$. Moreover,
complex Hadamard matrices found belong to the Butson class \cite{Butson1962,Tadej2006_2}
$\mathrm{BH}(d, \alpha)$, 
where
$d=pq, \ \alpha=1+\delta_{p,2}\delta_{r,1}$.

\item[(iii)] For $r\in\{1,2,\ldots,p-1\}$, the transition matrix
$V_{pq}=(H_{r}^{\mathrm{ent}})^\dagger H_{r+1}^{\mathrm{ent}}\in \mathcal{U}(pq)$ relating $H_r^{\mathrm{ent}}$ and
$H_{r+1}^{\mathrm{ent}}$ is specified by its action on the
computational basis as
\begin{equation}
V_{pq}\ket{s}
\sim
\sum_{t=0}^{pq-1}
\Bigg\{
\frac{1}{2\sqrt q}
\Big[
\tau_q^{-(s-t)^2}
+
\tau_{2q}^{(q+2)\left(2(s-t)+1\right)}
\tau_q^{-(s-t+1)^2}
\Big]\delta_{p,2}
+
\frac{1}{\sqrt{pq}}
\tau_p^{-(s-t)^2}
\tau_q^{-(s-t)^2}
(1-\delta_{p,2})
\Bigg\}\ket{t}.
\label{eq:Vfinal}
\end{equation}
Moreover, the transition matrix belongs to the Butson class \cite{Butson1962,Tadej2006_2}
$\mathrm{BH}(d, \alpha)$
 \ {\rm where}
$d=pq, \ \alpha=1+\delta_{p,2}$.
\end{enumerate} 
\end{thm}

\begin{proof}
We first prove (i). By direct computation, one can show that the MUBs introduced
in Theorem~\ref{thm:maxMUBs} are equivalent to those introduced
in Theorem~\ref{thm:startMUB}. To see this, we observe that for $r\in\{1,2,\ldots,p\}$
\begin{equation}
P_{pq}\ket{\chi(r,t)}
=\frac{1}{\sqrt{pq}}\sum_{j_p=0}^{p-1}\sum_{j_q=0}^{q-1}
\tau_p^{j_p[r(j_p-p)-2t]}
\tau_q^{j_q[r(j_q-q)-2t]}
\ket{j_p}\otimes
\ket{\eta_q(p)(j_q-j_p)\bmod q}
=\ket{\Psi(r,t)}.
\label{eq:perm_chi}
\end{equation} 
Such a result verifies the relation given in (\ref{eq:Hadamard_Relation}).
Since $P_{pq}$ is unitary and maps the bases in
Theorem \ref{thm:startMUB} onto those in
Theorem \ref{thm:maxMUBs}, the latter also form a set of
$p+1$ MUBs. 
The last equality in (\ref{eq:perm_chi}) follows from the fact that, for each fixed $j_p\in\mathbb{Z}_p$, any $j_q\in\mathbb{Z}_q$ can be uniquely expressed as $j_q\equiv j_p+pb\pmod q$, where $b\in\mathbb{Z}_q$. To simplify the expression, we first relabel the summation index $j_p$ as $a$ and introduce $n=a+pb$ for convenience. Next, we observe that
\begin{equation}
n[r(n-p)-2t]=a[r(a-p)-2t]+2pb(ra-t)+rp^2b(b-1).
\label{eq:equality}
\end{equation}
From (\ref{eq:equality}), it follows that
\begin{equation}
n[r(n-p)-2t]\equiv a[r(a-p)-2t]\pmod{(1+\delta_{p,2})p}.
\label{eq:simplifytau_p}
\end{equation}
Furthermore, since $q$ is odd we note that 
\begin{equation}
n[r(n-q)-2t]\equiv rn^2-2tn\pmod q.
\label{eq:simplify_tau_q}
\end{equation}
Substituting (\ref{eq:simplifytau_p}) and (\ref{eq:simplify_tau_q}) into the corresponding phase factors yields the desired expression.

We recall that any bipartite pure state
$\ket{\psi}_{AB}\in\mathcal{H}_A\otimes\mathcal{H}_B$ can be
expressed as
\begin{equation}
  \ket{\psi}_{AB}=
    \sum_{a=0}^{d_A-1}\sum_{b=0}^{d_B-1}
    C_{ab}\ket{a}_A\otimes\ket{b}_B,
\end{equation}
where $d_A$ ($d_B$) is the dimension of the Hilbert space $\mathcal{H}_A$ ($\mathcal{H}_B$) and $C\in\mathbb{C}^{d_A\times d_B}$ is the coefficient matrix of the state. If $d_A\leq d_B$, then the state $\ket{\psi}_{AB}$ is maximally entangled if and only if $CC^\dagger=\id/d_A$ \cite{Bengtsson2017}.

The coefficient matrix of the state $\ket{\Psi(r,t)}$ is given by
\begin{equation}
C_{ab}=\frac{1}{\sqrt{pq}}
\tau_{pq}^{(p+q)\left[
r(a+pb)^2-2t(a+pb)
\right]}
\end{equation}
for $r\in\{1,2,\ldots,p\}$.
By direct computation, we obtain
\begin{equation}
\begin{aligned}
(CC^\dagger)_{aa'}
&=
\frac{\tau_{pq}^{(p+q)(a-a')
\left[r(a+a')-2t\right]}}{pq}
G\left(0,r(p+q)(a-a'),q\right) \\
&=
\frac{\tau_{pq}^{(p+q)(a-a')
\left[r(a+a')-2t\right]}}{pq}
q\,\delta_{r(p+q)(a-a')\bmod q,0}
=\frac{1}{p}\delta_{aa'}.
\end{aligned}
\label{eq:reduced_matrix}
\end{equation}
The last equality follows from the fact that
$\delta_{r(p+q)(a-a')\bmod q,0}=\delta_{aa'}$
for $r\in\{1,2,\ldots,p\}$ and $p<q$.
This establishes that every vector in $\mathcal{B}_r^\mathrm{ent}$,
is maximally entangled for $r\in\{1,2,\ldots,p\}$. By construction, the states in $\mathcal{B}_0^\mathrm{sep}$
are separable. This completes the proof of (i).

We now prove (ii). We observe that the vectors $\ket{\Psi(r,t)}$ have equimodular entries for $r\in\{1,2,\ldots,p\}$. 
This observation combined with (i) implies that the collection $\{H_r^\mathrm{ent}\}_{r=1}^{p}$ consists of $p$ maximally entangled MUHMs.
The vectors $\ket{\Psi(r,t)}$ satisfy (\ref{eq:perm_chi}) for $r\in\{1,2,\ldots,p\}$. This observation combined with Theorem \ref{thm:startMUB} implies that the matrices $H_r^\mathrm{ent}$ are equivalent to $F_p\otimes F_q$. In order to determine the Butson classes of the Hadamard matrices, we consider two cases. To simplify the notation, we introduce $n=a+pb$.

\textbf{Case 1: $p$ and $q$ are odd.}
Since $p$ and $q$ are odd, $(p+q)/2$ is an integer. Consequently, we have
\begin{equation}
\begin{aligned}
\tau_{pq}^{(p+q)\left[
rn^2-2tn
\right]}
=
\omega_{pq}^{\eta_{pq}(2)(p+q)
\left[
rn^2-2tn
\right]}.
\end{aligned}
\end{equation}
This implies that $H_r^\mathrm{ent}\in\mathrm{BH}(pq,pq)$. Such a result is consistent with (ii).

\textbf{Case 2: $p=2$ and $q$ is odd.}
By expressing $\tau_{2q}$ in terms of $\omega_{4q}$ via the identity $\tau_{2q} = \omega_{4q}^{2q+1}$, the phase of the corresponding matrix entry can be rewritten as
\begin{equation}
\tau_{2q}^{(q+2)\left[rn^2-2tn\right]}
=
\omega_{4q}^{(2q+1)(q+2)\left[rn^2-2tn\right]}.
\end{equation}
This implies that in general $H_r^\mathrm{ent}\in\mathrm{BH}(2q,4q)$. Moreover, we have $H_r^\mathrm{ent}\in\mathrm{BH}(2q,2q)$ for $r=2$. Such a result is consistent with (ii). This completes the proof of (ii).

We now prove (iii). We note that the transition matrix can be expressed as
\cite{Horn2012}
\begin{equation}
\ket{\Psi(r+1,s)}
=
\sum_{t=0}^{pq-1}\bra{t}V_{pq}\ket{s}\ket{\Psi(r,t)}.
\end{equation}
Since every integer $n\in\mathbb{Z}_{pq}$ can be uniquely expressed
as $n\equiv a+pb \bmod (pq)$, where $a\in\mathbb{Z}_p$ and $b\in\mathbb{Z}_q$,
the double sum can be rewritten as a single sum over
$\mathbb{Z}_{pq}$. Furthermore, for the sake of convenience, we apply the change of variable
$n\mapsto -n \pmod{pq}$ to obtain
\begin{equation}
\bra{t}V_{pq}\ket{s}
=
\frac{1}{pq}
\sum_{n=0}^{pq-1}
\tau_{pq}^{(p+q)\left[
n^2+2(s-t)n
\right]}.
\label{eq:Ustart}
\end{equation}
Equation (\ref{eq:Ustart}) implies that the transition matrix is a Hadamard matrix, since all its entries are equimodular.
In order to determine a closed form for the transition matrix and its Butson class, we consider two cases. 

\textbf{Case 1: $p$ and $q$ are odd}. Analogously to the previous analysis, since $(p+q)/2$ is an integer, we get
\begin{equation}
\tau_{pq}^{(p+q)\left[
n^2+2(s-t)n
\right]}
=
\omega_{pq}^{\eta_{pq}(2)(p+q)
\left[
n^2+2(s-t)n
\right]}.
\end{equation}
This implies $V_{pq}\in\mathrm{BH}(pq,pq)$ as stated in (iii). We notice that (\ref{eq:Ustart}) may be expressed as
\begin{equation}
\begin{aligned}
\bra{t}V_{pq}\ket{s}
&=
\frac{1}{pq}
G\left(\eta_{pq}(2)(p+q),
(p+q)(s-t),
pq
\right)
\\
&=
\frac{1}{pq}
G\left(\eta_p(2),s-t,p\right)
G\left(\eta_q(2),s-t,q\right)
\\
&=
\frac{\varepsilon_p\varepsilon_q}{\sqrt{pq}}
\jacobi{\eta_p(2)}{p}
\jacobi{\eta_q(2)}{q}
\tau_p^{-(s-t)^2}
\tau_q^{-(s-t)^2}.
\end{aligned}
\label{eq:Podd}
\end{equation}
We see that (\ref{eq:Podd}) coincides with (\ref{eq:Vfinal}) up to a global phase for $p$ and $q$ that are odd. In the intermediate calculations, we use the congruences
$\eta_{pq}(2)\equiv\eta_p(2)\pmod p$ and
$\eta_{pq}(2)\equiv\eta_q(2)\pmod q$.

\textbf{Case 2: $p=2$ and $q$ is odd}. Using identity $\tau_{2q} = \omega_{4q}^{2q+1}$ as before, the phase simplifies to
\begin{equation}
\tau_{2q}^{(q+2)\left[
n^2+2(s-t)n
\right]}
=
\omega_{4q}^{(2q+1)(q+2)
\left[
n^2+2(s-t)n
\right]}.
\end{equation}
This implies that $V_{pq}\in\mathrm{BH}(2q,4q)$ as stated in (iii). For $p=2$, note that the sum over $n\in\mathbb{Z}_{2q}$ can be split
into sums over even and odd indices. Setting $n=2m$ and
$n=2m+1$, respectively, with $m\in\mathbb{Z}_q$, in (\ref{eq:Ustart}) gives
\begin{equation}
\begin{aligned}
\bra{t}V_{pq}\ket{s}
&=
\frac{1}{2q}
\Big[
G\left(2,2(s-t),q\right)
+
\tau_{2q}^{(q+2)\left(2(s-t)+1\right)}
G\left(2,2(s-t+1),q\right)
\Big]
\\
&=
\frac{\varepsilon_q}{2\sqrt q}
\jacobi{2}{q}
\Big[
\tau_q^{-(s-t)^2}
+
\tau_{2q}^{(q+2)\left(2(s-t)+1\right)}
\tau_q^{-(s-t+1)^2}
\Big].
\end{aligned}
\label{eq:Peven}
\end{equation}
We see that (\ref{eq:Peven}) coincides with (\ref{eq:Vfinal}) up to a global phase for $p=2$ and $q$ odd.

The diagonal unitary matrix $\Delta\in \mathcal{U}(pq)$
and permutation matrix $\Pi\in \mathcal{U}(pq)$
are defined by their action on the computational basis as
\begin{equation}
\Delta\ket{j}
=
\tau_p^{-j^2+2\delta_{p,2}j}
\tau_q^{-j^2}\ket{j},
\
\Pi\ket{j}
=
\ket{j(p+q)\bmod pq}.
\end{equation}
By direct computation, we obtain
\begin{equation}
\begin{aligned}
\Delta F_{pq}^\dagger\Pi\Delta\ket{s}
=\frac{
\tau_p^{-s^2+2\delta_{p,2}s}\tau_q^{-s^2}
}{\sqrt{pq}}
\sum_{t=0}^{pq-1}
\omega_{pq}^{t(p+q)s}
\tau_p^{-t^2+2\delta_{p,2}t}
\tau_q^{-t^2}\ket{t}
\sim V_{pq}\ket{s}.
\end{aligned}
\end{equation}
This implies that $V_{pq}\sim\Delta F_{pq}^\dagger\Pi\Delta$.
As a consequence, the transition matrix is equivalent
to the Fourier matrix $F_{pq}$. Moreover, recall that
$F_{d_1}\otimes F_{d_2}$ is equivalent to $F_{d_1d_2}$
whenever $\gcd(d_1,d_2)=1$ \cite{Tadej2006_1}.
Since $p$ and $q$ are distinct primes, it follows that
$F_p\otimes F_q$ is equivalent to $F_{pq}$.
Hence, $V_{pq}$ is equivalent to $F_p\otimes F_q$.
This completes the proof of (iii).
\end{proof}
Examples of sets of maximally entangled MUHMs constructed according to Theorem~\ref{thm:maxMUBs}, along with the transition matrices between consecutive Hadamard matrices, are presented in Appendix B for the four lowest-dimensional composite Hilbert spaces considered.

\section{Concluding remarks}
\label{sec:summary}

In this work, we have constructed $p$ maximally entangled MUBs in the bipartite Hilbert space $\mathcal{H}_p\otimes\mathcal{H}_q$, where $p$ and $q$ are distinct primes such that $p<q$. Together with a separable product basis, these bases form a set of $p+1$ MUBs. Our construction therefore establishes the lower bound for the number of maximally entangled MUBs in $\mathcal{H}_p\otimes\mathcal{H}_q$, which reads $M(p,q)\geq p$.

A key feature of our approach is that it yields explicit formulas for the basis vectors for arbitrary pairs of distinct prime dimensions, without requiring a separate search for suitable algebraic structures in each case.
Our method  provides a systematic procedure based on eigenvectors of Weyl--Heisenberg operators, the Chinese Remainder Theorem, and quadratic Gaussian sums. By applying a suitably chosen permutation 
matrix $P_{pq}$ 
to a set of product MUBs, we preserve mutual unbiasedness while transforming the product bases into one separable basis and $p$ bases consisting entirely of maximally entangled states. This yields a uniform closed-form construction   $H_r^\mathrm{ent}=P_{pq} H_r^\mathrm{sep}$,
applicable to every pair of distinct primes $p$ and $q$.

We have also investigated the complex Hadamard matrices associated with the constructed bases. The $p$ maximally entangled MUHMs are of Butson type. Each is equivalent to $F_p\otimes F_q$, and hence to the Fourier matrix $F_{pq}$, since $p$ and $q$ are coprime. Furthermore, we have derived closed-form expressions for the transition matrices between consecutive Hadamard matrices and determined their corresponding Butson--Hadamard classes. These results provide additional insights into the algebraic structure of the constructed MUHMs and the relations between them.

The extendability of the constructed sets of MUBs is a natural question. In the smallest case, $p=2$ and $q=3$, the resulting set of three MUBs is unextendible. Indeed, our construction is obtained by applying a common permutation to a triple of product MUBs. In \cite{McNulty2012} it has been proved that no triple of product MUBs in dimension six can be extended even by a single vector that is unbiased with respect to all three bases. Since a common unitary transformation preserves mutual unbiasedness and extendability, their result implies that the three MUBs constructed here cannot be extended by an additional MUB. For larger values of $p$ and $q$, determining whether the constructed sets of $p+1$ MUBs admit further extensions remains an open problem. 

Further directions include investigating whether the present closed-form approach can be extended to local dimensions that are prime powers or more general composite numbers, as well as to multipartite systems. Such extensions may help clarify which aspects of the construction rely essentially on the primality and coprimality of the local dimensions and which persist in more general settings. They may also shed light on the relationship between the algebraic structure of MUBs in composite dimensions and their entanglement properties.

\section{Acknowledgments}
It is a pleasure to thank Rafa{\l} Bistro\'n
for useful comments and valuable interactions. 
The authors acknowledge funding by the European Union under 
ERC Advanced Grant TAtypic, Project No. 101142236.

\appendix

\section{Chinese Remainder Theorem and Gaussian Sums}
This appendix provides the number-theoretic identities used in the
evaluation of the transition matrix (\ref{eq:Vfinal}) and reduced density matrix of the first subsystem (\ref{eq:reduced_matrix}). Throughout this appendix, we use the notation established
in the main text, with particular reference to
equations (\ref{eq:gauss_def})--(\ref{eq:modular_inverse}).

We begin by recalling the
Chinese Remainder Theorem, which will allow us to decompose quadratic
Gaussian sums with composite moduli into products of Gaussian sums with
coprime moduli. We then present the relevant properties and
evaluations of these sums.
\begin{thm*}[Chinese Remainder Theorem]\label{thm:CRT}
Let $d=\prod_{i=1}^n d_i$, where $d_1,d_2,\ldots,d_n$ are pairwise
coprime. The Chinese Remainder Theorem \cite{Rose1994} states that
the map
\begin{equation}
x\mapsto (x_1,x_2,\ldots,x_n),
\end{equation}
with $x_i=x\bmod d_i$, defines a ring isomorphism
\begin{equation}
\mathbb Z_d\cong
\mathbb Z_{d_1}\times \mathbb{Z}_{d_2}\times\cdots\times\mathbb Z_{d_n}.
\end{equation}
Its inverse is given by
\begin{equation}
(x_1,x_2,\ldots,x_n)\mapsto
\sum_{i=1}^n
\frac{d}{d_i}\,
\eta_{d_i}\left(\frac{d}{d_i}\right)x_i
\pmod d.
\end{equation}
\end{thm*}

We next collect the standard identities for quadratic Gaussian sums 
needed in the main body of this work. In particular, the following
lemma includes their evaluation for odd prime moduli.

\begin{lem*}\label{lem:gauss_sum}
Let  $a,b,c\in \mathbb{Z}$. Then, the following statements hold.
\begin{enumerate}
\item[(i)] For $c\geq1$, we have 
\begin{equation}
G(0,b,c)
=
c\,\delta_{0,b\bmod c}.
\end{equation}
\begin{equation}
G(1,0,c)
=
\frac{1}{2}\sqrt{c}(1+\ii)
\left(1+\e^{-\ii\pi c/2}\right)
=
\begin{cases}
\sqrt{c} & c\equiv1\pmod4,\\[1mm]
\ii\sqrt{c} & c\equiv3\pmod4,\\[1mm]
(1+\ii)\sqrt{c} & c\equiv0\pmod4,\\[1mm]
0 & c\equiv2\pmod4.
\end{cases}
\end{equation}
Moreover, if $p$ is an odd prime, such that $\gcd(a,p)=1$, we have
\begin{equation}
G(a,0,p)=\jacobi{a}{p}G(1,0,p).
\end{equation}

\item[(ii)] Let $p$ be an odd prime,
such that $\operatorname{gcd}(a,p)=1$. Define $\psi_p(a)=\eta_p(4a)$. Then, we have
\begin{equation}
G(a,b,p)
=
\varepsilon_p\sqrt{p}\,
\jacobi{a}{p}\,
\omega_p^{-\psi_p(a)b^2}.
\end{equation}
\end{enumerate}
\end{lem*}

\begin{proof}
As (i) is proven in \cite{Apostol1976}, we focus our attention on proving (ii). 
We complete the square in the exponent and obtain
\begin{equation}
\begin{aligned}
an^2+bn
&\equiv
a\left(n+2\psi_p(a)b\right)^2
-\psi_p(a)b^2
\pmod p,
\end{aligned}
\end{equation}
In the computation, we used the congruence $4a\psi_p(a)\equiv1\pmod p$. This yields
\begin{equation}
\begin{aligned}
G(a,b,p)=
\omega_p^{-\psi_p(a)b^2}
\sum_{n=0}^{p-1}
\omega_p^{a\left(n+2\psi_p(a)b\right)^2}=
\omega_p^{-\psi_p(a)b^2}G(a,0,p),
\end{aligned}
\end{equation}
where the last equality follows from the fact that
$n\mapsto n+2\psi_p(a)b \bmod p$ is a permutation
of $\mathbb{Z}_p$.
Finally, using (i) and the fact that $p$ is odd, we obtain
\begin{equation}
G(a,0,p)
=
\varepsilon_p\sqrt{p}\jacobi{a}{p}.
\end{equation}
Result (ii) now follows directly.
\end{proof}

The preceding lemma provides the required evaluation of quadratic
Gaussian sums for odd prime moduli. The following lemma shows how a
Gaussian sum with modulus $pq$, where $p$ and $q$ are distinct primes,
can be factorized into a product of Gaussian sums with moduli $p$ and $q$.

\begin{lem*}\label{lem:gauss_sum_product}
Let $p$ and $q$ be distinct prime numbers.
Then, we have
\begin{equation}
G(a,b,pq)
=
G(a\eta_p(q),b\eta_p(q),p)
G(a\eta_q(p),b\eta_q(p),q).
\end{equation}
\end{lem*}

\begin{proof}
By the Chinese Remainder Theorem, every
$n\in\mathbb{Z}_{pq}$ can be uniquely represented as
\begin{equation}
n\equiv
q\eta_p(q)n_p
+
p\eta_q(p)n_q
\pmod{pq},
\end{equation}
where $n_p\in\mathbb{Z}_p$ and
$n_q\in\mathbb{Z}_q$. Thus,
\begin{equation}
\begin{aligned}
an^2+bn
\equiv
aq^2\eta_p(q)^2n_p^2
+
bq\eta_p(q)n_p
+
ap^2\eta_q(p)^2n_q^2
+
bp\eta_q(p)n_q
+
2apq\eta_p(q)\eta_q(p)n_pn_q
\pmod{pq}.
\end{aligned}
\end{equation}
The mixed term vanishes modulo $pq$, since $2apq\eta_p(q)\eta_q(p)n_pn_q
\equiv0\pmod{pq}$. The remaining terms can be rearranged using $q\eta_p(q)\equiv1\pmod p$ and $p\eta_q(p)\equiv1\pmod q$. Consequently, we obtain
\begin{equation}
an^2+bn
\equiv
q\eta_p(q)(an_p^2+bn_p)
+
p\eta_q(p)(an_q^2+bn_q)
\pmod{pq}.
\end{equation}
Finally, by the Chinese Remainder Theorem, the sum over
$n\in\mathbb{Z}_{pq}$ can be rewritten as a sum over pairs
$(n_p,n_q)\in\mathbb{Z}_p\times\mathbb{Z}_q$.
Using $\omega_{pq}^{q}=\omega_p$ and
$\omega_{pq}^{p}=\omega_q$, the double sum
factorizes into the product of two Gaussian sums, yielding the
sought-after result.
\end{proof}

\section{Examplary unbiased complex Hadamard matrices of order $d=pq$}
\label{app:examples}

We present examples of sets of $p$ maximally entangled MUHMs in
$\mathcal{H}_p\otimes\mathcal{H}_q$, where $p<q$. They are constructed
according to Theorem \ref{thm:maxMUBs}
by applying the permutation matrix $P_{pq}$
on the standard solution of separable MUBs. 
For each of the four
lowest-dimensional composite Hilbert spaces considered, we provide the corresponding
Hadamard matrices $H_r^\mathrm{ent}=P_{pq} H_r^\mathrm{sep}\in
\mathrm{BH}\!\left(
pq,\,\alpha
\right)$
with $\alpha=pq( 1+\delta_{p,2}\delta_{r,1})$
as well as the transition matrices $V_{pq}\in\mathrm{BH}\bigl(pq,
\alpha)$ with $\alpha=pq(1+\delta_{p,2})$. 
The transition matrices $V_{pq}$, relating consecutive Hadamard matrices,
$H_{r+1}^\mathrm{ent}=H_r^\mathrm{ent} V_{pq}$, 
are not of the product form, 
$V_{pq} \ne V_p \otimes V_q$. This observation follows directly from the operator Schmidt decomposition \cite{Bengtsson2017}.
For simplicity,
all matrices are presented without normalization factors, up to
a global phase and column dephasing. Following the notation of
\cite{Tadej2006_2}, $\operatorname{LOG}(\cdot)$ denotes the entrywise
logarithm, and $\bullet$ represents a zero entry. The $t$-th
column of $H_r^\mathrm{ent}$ corresponds to the vector $\ket{\Psi(r,t)}$.


\subsection*{Dimension six, $p=2, q=3$}


\begin{align}
\operatorname{LOG}(H_1^\mathrm{ent})=\frac{2\pi\ii}{12}
\begin{pmatrix}
\bullet&\bullet&\bullet&\bullet&\bullet&\bullet\\
8&\bullet&4&8&\bullet&4\\
8&4&\bullet&8&4&\bullet\\
11&1&3&5&7&9\\
3&9&3&9&3&9\\
11&9&7&5&3&1
\end{pmatrix},\
\operatorname{LOG}(H_2^\mathrm{ent})=\frac{2\pi\ii}{6}
\begin{pmatrix}
\bullet&\bullet&\bullet&\bullet&\bullet&\bullet\\
2&4&\bullet&2&4&\bullet\\
2&\bullet&4&2&\bullet&4\\
5&\bullet&1&2&3&4\\
3&\bullet&3&\bullet&3&\bullet\\
5&4&3&2&1&\bullet
\end{pmatrix}, \
\operatorname{LOG}(V_6)=\frac{2\pi\ii}{12}
\begin{pmatrix}
\bullet&\bullet&\bullet&\bullet&\bullet&\bullet\\
1&11&9&7&5&3\\
4&\bullet&8&4&\bullet&8\\
9&3&9&3&9&3\\
4&8&\bullet&4&8&\bullet\\
1&3&5&7&9&11
\end{pmatrix}.
\end{align}


\subsection*{Dimension ten, $p=2, q=5$}


\begin{align}
&\operatorname{LOG}(H_1^\mathrm{ent})=\frac{2\pi\ii}{20}
\begin{pmatrix}
\bullet&\bullet&\bullet&\bullet&\bullet&\bullet&\bullet&\bullet&\bullet&\bullet\\
8&\bullet&12&4&16&8&\bullet&12&4&16\\
12&16&\bullet&4&8&12&16&\bullet&4&8\\
12&8&4&\bullet&16&12&8&4&\bullet&16\\
8&16&4&12&\bullet&8&16&4&12&\bullet\\
17&3&9&15&1&7&13&19&5&11\\
13&11&9&7&5&3&1&19&17&15\\
5&15&5&15&5&15&5&15&5&15\\
13&15&17&19&1&3&5&7&9&11\\
17&11&5&19&13&7&1&15&9&3
\end{pmatrix}, \\
&\operatorname{LOG}(H_2^\mathrm{ent})=\frac{2\pi\ii}{10}
\begin{pmatrix}
\bullet&\bullet&\bullet&\bullet&\bullet&\bullet&\bullet&\bullet&\bullet&\bullet\\
8&4&\bullet&6&2&8&4&\bullet&6&2\\
2&4&6&8&\bullet&2&4&6&8&\bullet\\
2&\bullet&8&6&4&2&\bullet&8&6&4\\
8&2&6&\bullet&4&8&2&6&\bullet&4\\
7&\bullet&3&6&9&2&5&8&1&4\\
3&2&1&\bullet&9&8&7&6&5&4\\
5&\bullet&5&\bullet&5&\bullet&5&\bullet&5&\bullet\\
3&4&5&6&7&8&9&\bullet&1&2\\
7&4&1&8&5&2&9&6&3&\bullet
\end{pmatrix}, \\
&\operatorname{LOG}(V_{10})=\frac{2\pi\ii}{20}
\begin{pmatrix}
\bullet&\bullet&\bullet&\bullet&\bullet&\bullet&\bullet&\bullet&\bullet&\bullet\\
3&17&11&5&19&13&7&1&15&9\\
12&\bullet&8&16&4&12&\bullet&8&16&4\\
7&9&11&13&15&17&19&1&3&5\\
8&4&\bullet&16&12&8&4&\bullet&16&12\\
15&5&15&5&15&5&15&5&15&5\\
8&12&16&\bullet&4&8&12&16&\bullet&4\\
7&5&3&1&19&17&15&13&11&9\\
12&4&16&8&\bullet&12&4&16&8&\bullet\\
3&9&15&1&7&13&19&5&11&17
\end{pmatrix}.
\end{align}


\subsection*{Dimension fourteen, $p=2, q=7$}

\begin{align}
&\operatorname{LOG}(H_1^\mathrm{ent})=\frac{2\pi\ii}{28}
\begin{pmatrix}
\bullet&\bullet&\bullet&\bullet&\bullet&\bullet&\bullet&\bullet&\bullet&\bullet&\bullet&\bullet&\bullet&\bullet\\
8&\bullet&20&12&4&24&16&8&\bullet&20&12&4&24&16\\
4&16&\bullet&12&24&8&20&4&16&\bullet&12&24&8&20\\
16&20&24&\bullet&4&8&12&16&20&24&\bullet&4&8&12\\
16&12&8&4&\bullet&24&20&16&12&8&4&\bullet&24&20\\
4&20&8&24&12&\bullet&16&4&20&8&24&12&\bullet&16\\
8&16&24&4&12&20&\bullet&8&16&24&4&12&20&\bullet\\
23&5&15&25&7&17&27&9&19&1&11&21&3&13\\
11&13&15&17&19&21&23&25&27&1&3&5&7&9\\
15&9&3&25&19&13&7&1&23&17&11&5&27&21\\
7&21&7&21&7&21&7&21&7&21&7&21&7&21\\
15&21&27&5&11&17&23&1&7&13&19&25&3&9\\
11&9&7&5&3&1&27&25&23&21&19&17&15&13\\
23&13&3&21&11&1&19&9&27&17&7&25&15&5
\end{pmatrix}, \\
&\operatorname{LOG}(H_2^\mathrm{ent})=\frac{2\pi\ii}{14}
\begin{pmatrix}
\bullet&\bullet&\bullet&\bullet&\bullet&\bullet&\bullet&\bullet&\bullet&\bullet&\bullet&\bullet&\bullet&\bullet\\
8&4&\bullet&10&6&2&12&8&4&\bullet&10&6&2&12\\
4&10&2&8&\bullet&6&12&4&10&2&8&\bullet&6&12\\
2&4&6&8&10&12&\bullet&2&4&6&8&10&12&\bullet\\
2&\bullet&12&10&8&6&4&2&\bullet&12&10&8&6&4\\
4&12&6&\bullet&8&2&10&4&12&6&\bullet&8&2&10\\
8&12&2&6&10&\bullet&4&8&12&2&6&10&\bullet&4\\
9&\bullet&5&10&1&6&11&2&7&12&3&8&13&4\\
11&12&13&\bullet&1&2&3&4&5&6&7&8&9&10\\
1&12&9&6&3&\bullet&11&8&5&2&13&10&7&4\\
7&\bullet&7&\bullet&7&\bullet&7&\bullet&7&\bullet&7&\bullet&7&\bullet\\
1&4&7&10&13&2&5&8&11&\bullet&3&6&9&12\\
11&10&9&8&7&6&5&4&3&2&1&\bullet&13&12\\
9&4&13&8&3&12&7&2&11&6&1&10&5&\bullet
\end{pmatrix}, \\
&\operatorname{LOG}(V_{14})=\frac{2\pi\ii}{28}
\begin{pmatrix}
\bullet&\bullet&\bullet&\bullet&\bullet&\bullet&\bullet&\bullet&\bullet&\bullet&\bullet&\bullet&\bullet&\bullet\\
5&23&13&3&21&11&1&19&9&27&17&7&25&15\\
20&\bullet&8&16&24&4&12&20&\bullet&8&16&24&4&12\\
17&15&13&11&9&7&5&3&1&27&25&23&21&19\\
24&12&\bullet&16&4&20&8&24&12&\bullet&16&4&20&8\\
13&19&25&3&9&15&21&27&5&11&17&23&1&7\\
12&8&4&\bullet&24&20&16&12&8&4&\bullet&24&20&16\\
21&7&21&7&21&7&21&7&21&7&21&7&21&7\\
12&16&20&24&\bullet&4&8&12&16&20&24&\bullet&4&8\\
13&7&1&23&17&11&5&27&21&15&9&3&25&19\\
24&8&20&4&16&\bullet&12&24&8&20&4&16&\bullet&12\\
17&19&21&23&25&27&1&3&5&7&9&11&13&15\\
20&12&4&24&16&8&\bullet&20&12&4&24&16&8&\bullet\\
5&15&25&7&17&27&9&19&1&11&21&3&13&23
\end{pmatrix}.
\end{align}

\subsection*{Dimension fifteen, $p=3, q=5$}
%
%
\begin{align}
&\operatorname{LOG}(H_1^\mathrm{ent})=\frac{2\pi\ii}{15}\begin{pmatrix}
\bullet&\bullet&\bullet&\bullet&\bullet&\bullet&\bullet&\bullet&\bullet&\bullet&\bullet&\bullet&\bullet&\bullet&\bullet\\
6&12&3&9&\bullet&6&12&3&9&\bullet&6&12&3&9&\bullet\\
9&6&3&\bullet&12&9&6&3&\bullet&12&9&6&3&\bullet&12\\
9&12&\bullet&3&6&9&12&\bullet&3&6&9&12&\bullet&3&6\\
6&\bullet&9&3&12&6&\bullet&9&3&12&6&\bullet&9&3&12\\
4&11&3&10&2&9&1&8&\bullet&7&14&6&13&5&12\\
4&2&\bullet&13&11&9&7&5&3&1&14&12&10&8&6\\
1&5&9&13&2&6&10&14&3&7&11&\bullet&4&8&12\\
10&5&\bullet&10&5&\bullet&10&5&\bullet&10&5&\bullet&10&5&\bullet\\
1&2&3&4&5&6&7&8&9&10&11&12&13&14&\bullet\\
1&\bullet&14&13&12&11&10&9&8&7&6&5&4&3&2\\
10&\bullet&5&10&\bullet&5&10&\bullet&5&10&\bullet&5&10&\bullet&5\\
1&12&8&4&\bullet&11&7&3&14&10&6&2&13&9&5\\
4&6&8&10&12&14&1&3&5&7&9&11&13&\bullet&2\\
4&12&5&13&6&14&7&\bullet&8&1&9&2&10&3&11
\end{pmatrix}, \\
&\operatorname{LOG}(H_2^\mathrm{ent})=\frac{2\pi\ii}{15}
\begin{pmatrix}
\bullet&\bullet&\bullet&\bullet&\bullet&\bullet&\bullet&\bullet&\bullet&\bullet&\bullet&\bullet&\bullet&\bullet&\bullet\\
12&3&9&\bullet&6&12&3&9&\bullet&6&12&3&9&\bullet&6\\
3&\bullet&12&9&6&3&\bullet&12&9&6&3&\bullet&12&9&6\\
3&6&9&12&\bullet&3&6&9&12&\bullet&3&6&9&12&\bullet\\
12&6&\bullet&9&3&12&6&\bullet&9&3&12&6&\bullet&9&3\\
8&\bullet&7&14&6&13&5&12&4&11&3&10&2&9&1\\
8&6&4&2&\bullet&13&11&9&7&5&3&1&14&12&10\\
2&6&10&14&3&7&11&\bullet&4&8&12&1&5&9&13\\
5&\bullet&10&5&\bullet&10&5&\bullet&10&5&\bullet&10&5&\bullet&10\\
2&3&4&5&6&7&8&9&10&11&12&13&14&\bullet&1\\
2&1&\bullet&14&13&12&11&10&9&8&7&6&5&4&3\\
5&10&\bullet&5&10&\bullet&5&10&\bullet&5&10&\bullet&5&10&\bullet\\
2&13&9&5&1&12&8&4&\bullet&11&7&3&14&10&6\\
8&10&12&14&1&3&5&7&9&11&13&\bullet&2&4&6\\
8&1&9&2&10&3&11&4&12&5&13&6&14&7&\bullet
\end{pmatrix}, \\
&\operatorname{LOG}(H_3^\mathrm{ent})=\frac{2\pi\ii}{15}
\begin{pmatrix}
\bullet&\bullet&\bullet&\bullet&\bullet&\bullet&\bullet&\bullet&\bullet&\bullet&\bullet&\bullet&\bullet&\bullet&\bullet\\
3&9&\bullet&6&12&3&9&\bullet&6&12&3&9&\bullet&6&12\\
12&9&6&3&\bullet&12&9&6&3&\bullet&12&9&6&3&\bullet\\
12&\bullet&3&6&9&12&\bullet&3&6&9&12&\bullet&3&6&9\\
3&12&6&\bullet&9&3&12&6&\bullet&9&3&12&6&\bullet&9\\
12&4&11&3&10&2&9&1&8&\bullet&7&14&6&13&5\\
12&10&8&6&4&2&\bullet&13&11&9&7&5&3&1&14\\
3&7&11&\bullet&4&8&12&1&5&9&13&2&6&10&14\\
\bullet&10&5&\bullet&10&5&\bullet&10&5&\bullet&10&5&\bullet&10&5\\
3&4&5&6&7&8&9&10&11&12&13&14&\bullet&1&2\\
3&2&1&\bullet&14&13&12&11&10&9&8&7&6&5&4\\
\bullet&5&10&\bullet&5&10&\bullet&5&10&\bullet&5&10&\bullet&5&10\\
3&14&10&6&2&13&9&5&1&12&8&4&\bullet&11&7\\
12&14&1&3&5&7&9&11&13&\bullet&2&4&6&8&10\\
12&5&13&6&14&7&\bullet&8&1&9&2&10&3&11&4
\end{pmatrix}, \\
&\operatorname{LOG}(V_{15})=\frac{2\pi\ii}{15}
\begin{pmatrix}
\bullet&\bullet&\bullet&\bullet&\bullet&\bullet&\bullet&\bullet&\bullet&\bullet&\bullet&\bullet&\bullet&\bullet&\bullet\\
11&4&12&5&13&6&14&7&\bullet&8&1&9&2&10&3\\
14&\bullet&1&2&3&4&5&6&7&8&9&10&11&12&13\\
9&3&12&6&\bullet&9&3&12&6&\bullet&9&3&12&6&\bullet\\
11&13&\bullet&2&4&6&8&10&12&14&1&3&5&7&9\\
5&\bullet&10&5&\bullet&10&5&\bullet&10&5&\bullet&10&5&\bullet&10\\
6&9&12&\bullet&3&6&9&12&\bullet&3&6&9&12&\bullet&3\\
14&10&6&2&13&9&5&1&12&8&4&\bullet&11&7&3\\
14&3&7&11&\bullet&4&8&12&1&5&9&13&2&6&10\\
6&3&\bullet&12&9&6&3&\bullet&12&9&6&3&\bullet&12&9\\
5&10&\bullet&5&10&\bullet&5&10&\bullet&5&10&\bullet&5&10&\bullet\\
11&9&7&5&3&1&14&12&10&8&6&4&2&\bullet&13\\
9&\bullet&6&12&3&9&\bullet&6&12&3&9&\bullet&6&12&3\\
14&13&12&11&10&9&8&7&6&5&4&3&2&1&\bullet\\
11&3&10&2&9&1&8&\bullet&7&14&6&13&5&12&4
\end{pmatrix}.
\end{align}

\bibliography{references.bib}

\end{document}